%% file: main.tex
\documentclass{article}

\PassOptionsToPackage{numbers,sort&compress}{natbib}

\usepackage[main, final]{iaseai26}

\usepackage[utf8]{inputenc} 
\usepackage[T1]{fontenc}    
\usepackage{hyperref}       
\usepackage{url}            
\usepackage{booktabs}       
\usepackage{amsfonts}       
\usepackage{nicefrac}       
\usepackage{microtype}      
\usepackage{xcolor}         
\usepackage{tabularx}
\usepackage{array}

\title{The Agony of Opacity:\\ Foundations for Reflective Interpretability in AI-Mediated Mental Health Support}

\author{
{\normalfont Sachin R. Pendse\textsuperscript{1}}
\and Darren Gergle\textsuperscript{2}
\and Rachel Kornfield\textsuperscript{3}
\and Kaylee Kruzan\textsuperscript{3}
\and David Mohr\textsuperscript{3}
\and Jessica Schleider\textsuperscript{3}
\and Jina Suh\textsuperscript{4}
\and Annie Wescott\textsuperscript{3}
\and Jonah Meyerhoff\textsuperscript{3}
\\[0.75ex]
\textsuperscript{1}University of California, San Francisco (UCSF), San Francisco, United States\\
\textsuperscript{2}Northwestern University, Evanston, United States\\
\textsuperscript{3}Feinberg School of Medicine, Northwestern University, Chicago, United States\\
\textsuperscript{4}Microsoft Research, Redmond, United States\\[0.75ex]
\texttt{Corresponding author: Sachin Pendse (sachin.pendse@ucsf.edu)}
}

\begin{document}
\maketitle
\vspace{-15pt}
\begin{abstract}
\vspace{-8pt}
\input{new_sections/0_Abstract.tex}
\end{abstract}

\section{Introduction}
\input{new_sections/1_Introduction.tex}

\section{Psychological Distress States and Interaction Patterns with AI-Mediated Technologies}
\input{new_sections/2_Psychological_Distress.tex}

\section{Building Reflective Interpretation into AI-Mediated Mental Health Support: The Ethical Case for a Higher Standard of Interpretability}
\input{new_sections/3_Building_Interpretability.tex}

\section{Reflective Interpretability Strategies from Mental Health Research Fields}
\input{new_sections/4_Interpretability_Strategies.tex}

\section{Tensions and Potential Risks}
\input{new_sections/5_Policy_Implications.tex}

\section{Conclusion}
\input{new_sections/6_Conclusion.tex}

\pagebreak
\appendix
\section*{Appendix A: Key Insights for Reflective Interpretability}
In this appendix, we provide core takeaways and key insights from this paper.

\medskip 
\noindent\begin{tabularx}{\textwidth}{@{}>{\raggedright\arraybackslash}p{0.40\textwidth} X@{}}
\toprule
\textbf{Question} & \textbf{Key Takeaway} \\
\midrule
\textit{What is reflective interpretability?} &  We understand reflective interpretability to be an agency-preserving and iterative process of reflecting on model outputs related to one's experience of distress, and interpreting that information towards creating meaning from it. We delineate three core aspects of reflective interpretability, occurring throughout engagement with AI-mediated mental health support: 1) the ability of users to actively make sense of how model outputs were generated, 2) clarity on the fixed boundaries of support that can be provided by a model (including what kinds of interactions might trigger those boundaries), and 3) functionality that encourages users to reflect on model outputs and interpret outputs in ways that promote long-term well-being \\ \addlinespace[1.2em]
\textit{Why is it important to take into consideration in AI-mediated mental health support?} &  Past research has demonstrated how the acute experience of distress can alter an individual's perspective in ways that make opacity particularly dangerous. A state of acute distress may make it easier to accept new information as fact, and act on it as directive advice, particularly if it appears to be coming from an authority figure (as is often the case with AI chatbot interfaces). \\ \addlinespace[1.2em]
\textit{What are the conceptual precedents of reflective interpretability?} & The conceptual precedents of reflective interpretability originate in research from medical ethics around informed consent being a reflective and interpretative dialogue between provider and client, centered on active discussion of needs and boundaries, alongside research from HCI around interface design that centers user agency and authority (rather than that of the technical system). \\ \addlinespace[1.2em]
\textit{Why do we call it reflective \textit{interpretability}?} & The practice of mental health support fundamentally resists providing objective explanations of behavior or experience to individuals, and encourages client reflection and \textit{interpretation}, with the client acting as an active expert in their healing process.\\ \addlinespace[1.2em]
\textit{How might reflective interpretability be integrated into the interfaces associated with AI-mediated mental health support?} & Role induction (from psychotherapy), prosocial advance directives (from crisis intervention), intervention titration (from psychiatry), and mechanisms for recourse (from care authorization). \\ \addlinespace[1.2em]
\textit{What potential tensions or harms exist alongside a reflective interpretability approach? And how might these be mitigated?} & Adding more reflective interpretability features to an interface might cause additional friction in an environment where friction might disincentivize engaging with (potentially life-saving) support. In addition, showing too much of the model's internal reasoning could be perceived as manipulative or offensive. More research is needed to understand how best to integrate these features such that they do not make getting support \textit{more} difficult. \\
\bottomrule
\end{tabularx}

{\small
\bibliographystyle{unsrtnat} 
\bibliography{references}     
}

\end{document}

%% file: new_sections/0_Abstract.tex
Throughout history, a prevailing paradigm in mental healthcare has been one in which distressed people may receive treatment with little understanding around how their experience is perceived by their care provider, and in turn, the decisions made by their provider around how treatment will progress. Paralleling this offline model of care, people who seek mental health support from artificial intelligence (AI)-based chatbots are similarly provided little context for how their expressions of distress are processed by the model, and subsequently, any reasoning or theoretical grounding that may underlie model responses. People in severe distress who turn to AI chatbots for support thus find themselves caught between black boxes, contending with unique forms of agony that arise from these intersecting opacities. In this paper, we argue that the distinct psychological state of individuals experiencing severe mental distress uniquely necessitates a higher standard of end-user interpretability in comparison to general AI chatbot use. We propose a \textit{reflective interpretability} approach to AI-mediated mental health support, which nudges users to engage in an agency-preserving and iterative process of reflection and interpretation of model outputs, towards creating meaning from interactions (rather than accepting outputs as directive instructions). Drawing on interpretability practices from four mental health fields (psychotherapy, crisis intervention, psychiatry, and care authorization), we describe concrete design approaches for reflective interpretability in AI-mediated mental health support, including role induction, prosocial advance directives, intervention titration, and well-defined mechanisms for recourse, alongside a discussion of potential risks and mitigation measures. 

%% file: new_sections/1_Introduction.tex
For an individual in distress, across geographic contexts, the search for mental healthcare is punctuated by repeated interactions with opaque black boxes. During a visit with a general practitioner, individuals may be asked sensitive questions about whether they are experiencing suicidal ideation without fully understanding how the information they provide might be used~\cite{richards2019understanding}, or during a psychotherapy session, clients may feel a sense of mystery~\cite{turns2019removing} about why certain questions are being asked, to the point of feeling as if they ``[don't] understand the rules [of the game]''~\cite{vybiral2024negative}. Patients may leave a psychiatric appointment with a medication prescription, but little sense of how that medication might work for them~\cite{luitel2020perception, teferra2013perspectives}, and looking to crisis intervention, clients may be involuntarily hospitalized with little insight into why~\cite{akther2019patients, wong2020experiences}. After treatment has been administered, patients may be surprised to receive a bill from their healthcare provider for services they did not realize they received (such as being billed for ``outpatient observation'' during an emergency psychiatric consultation~\cite{kangovi2015patient}), or in countries with centralized or publicly funded health systems, they may find that they owe hospital, prescription, or maintenance fees after their treatment~\cite{ameli_forfait_hospitalier_2025, citytoronto_ambulance_fees, stjohn_ambulance_charges_2025}. In practice, these black boxes prevent individuals from understanding the reasoning behind their care or picturing their path to recovery, which may increase distress~\cite{vybiral2024negative}. The use of artificial intelligence (AI)-based chatbots for mental health support can add a new black box to this process, with similar implications. 

AI-mediated mental health support can be quite helpful for those experiencing acute distress~\cite{song2024typing, siddals2024happened, petersson2025believe, jung2025ve}. However, interfaces can mirror the black box nature of offline mental health support, with users confused about where responses originate or why certain responses are provided. Interview studies of AI chatbot mental health support seekers~\cite{song2024typing, siddals2024happened, petersson2025believe} have described little understanding of how responses are produced. The harms of these opaque AI systems are well-discussed in explainable AI and AI safety literature~\cite{akbulut2024all}, in ways that parallel how the harms of a lack of psychoeducation are discussed in mental healthcare contexts~\cite{motlova2017psychoeducation}. However, these two opacities intersect and compound in a unique way when people in distress use AI chatbots for mental health support, with documented real-world harms. Lacking sufficient information about how outputs are constructed within a seemingly authoritative interface, users may treat responses as prescriptive or originating from an authority figure, and neglect to engage in a nuanced process of \textit{interpreting} outputs given potential model biases or limitations (unlike the reflective interpretation they might engage in within psychotherapy~\cite{sidis2023conceptualisation}). 

For an individual experiencing the profound vulnerability and isolation of mental distress, this opacity, nested within an authoritative interface, can lead to further agonizing situations. For instance, Hill~\cite{hill_chatgpt_conspiracies_2025}, Klee~\cite{klee2025ozarks}, and Jargon~\cite{jargon_chatgpt_mania_2025} describe incidents in which users began to lean on AI chatbots for support, but increasingly understood chatbots as being sentient, and became deeply distressed about the chatbot's welfare as they experienced beliefs of revolutionary scientific discoveries, reinforced by chatbot responses. Klee~\cite{klee_bodydysmorphia_ai_2025} similarly describes how people with body dysmorphia can ask AI chatbots for ``reassurance and `objective' evaluations of attractiveness,'' which in many cases, provide results that are ``viciously insulting, [and] not the sort of thing anyone would want to read about themselves''~\cite{klee_bodydysmorphia_ai_2025}. Roose describes how an AI chatbot told a teenage user who was suicidal to ``come home,'' with the user taking his life immediately after~\cite{roose_ai_suicide_2024}. Clear real-world harms originate from users misinterpreting model outputs, and these issues take on particular relevance when considering the vulnerable psychological state of people experiencing severe mental distress.

\textbf{Statement of Contributions:} In this paper, we argue that ensuring the safety of AI-mediated mental health support necessitates a more expansive understanding of interpretability, viewed from the end-user's perspective and grounded in the potential for interface design to ensure that users engage in a reflective process of interpreting model outputs. We call this process \textit{reflective interpretability}, grounding our argument for its importance in empirical clinical research on the altered state of mind of someone experiencing distress, and how that state of mind influences technology perceptions and use. Drawing on reflective interpretability practices from four mental health domains, we propose concrete technical foundations for AI-mediated mental health support that encourages reflection and interpretation. As part of each foundation, we also introduce literature from medical ethics (around interpretability as a consent-driven \textit{dialogue}) into the emerging domain of AI safety within mental health contexts. We then conclude with a discussion of the potential tensions and risks associated with reflective interpretability, and the role of design and policy towards mitigating these risks.

%% file: new_sections/2_Psychological_Distress.tex
Below, we describe how use of AI-mediated mental health support often happens in moments of dire need, with little of the educational scaffolding and transparency that typically accompanies mental health services. We argue that, given the unique state of mind of an individual in distress, this lack of scaffolding may be a primary factor in documented real-world harms.  

\subsection{Current Usage of AI-Mediated Mental Health Support}
Traditional mental health support typically begins with information around how support will be provided and why it might be helpful, often called anticipatory socialization~\cite{orne1968anticipatory} or role induction~\cite{swift2023meta}. This process can include aligning an individual's expectations to the kinds of support that can be provided and establishing clear boundaries~\cite{gray2013introduction, swift2023meta, orne1968anticipatory}. For instance, this may include clarifying the role of the supporter and how it can be distinguished from other relationships (such as platonic or romantic relationships), explaining the limits of confidentiality before any sensitive information is disclosed, describing the specific therapeutic approach and the rationale for how interactions within session will be structured, and describing pathways for recourse if an individual does not feel supported.

Past work~\cite{song2024typing, siddals2024happened, petersson2025believe, jung2025ve} has demonstrated that users' points of first engagement with AI-mediated mental health support largely \textit{lack} role induction, with users turning to AI chatbots due to acute mental health need and a familiarity with empathetic AI chatbot interfaces. For instance, Song et al.~\cite{song2024typing} describe how none of the participants they interviewed first used AI chatbots for mental health support, with several participants describing how initial uses were for computer programming assistance. Similarly, both Jung et al.~\cite{jung2025ve} and Siddals et al.~\cite{siddals2024happened} describe how users primarily turned to AI chatbots in moments of intense distress and when no other supports were available or financially accessible. Some participants also held the support that was provided by mental health chatbots in high regard, owing to the authoritative nature of the interface. For instance, Song et al.~\cite{song2024typing} describe how participants would use information from the bots to validate psychiatric diagnoses or diagnose others, Siddals et al.~\cite{siddals2024happened} describe how one participant noted ``\textit{it’s pure science...ChatGPT is telling me what is correct to do},'' and Petersson et al.~\cite{petersson2025believe} describe participant beliefs that AI chatbot systems will soon detect mental health issues before the user is even aware of them.

This lack of role induction within current chatbot applications may lower barriers to engagement by allowing users to engage with support in ways that feel informal and low-stakes. Past work has demonstrated that in offline~\cite{adams2025administrative, coombs2021barriers, barrett2008early} and online~\cite{smith2025engagement, jardine2024between} contexts, increasing the number of barriers to care can increase attrition rates. However, given that users report being persuaded to seek support due to the friendly and conversational interface associated with AI chatbots~\cite{song2024typing, jung2025ve, morrin2025delusions, yeung2025psychogenic}, many users mix mundane and everyday needs with mental health support needs, without being provided the clear legal and ethical boundaries typically provided explicitly in traditional mental healthcare contexts. Given the vulnerable state of mind of people in distress, this lack of clear scaffolding may create situations where users receive inadequate or harmful advice without the protective interpretability mechanisms that underlie engagement with traditional mental health services (such as clarifying concerns with a provider). Documented real-world harms from engagements with AI-mediated mental health support speak to the danger of this intersection between the unique vulnerabilities associated with a distressed state of mind, and the authority attributed to interfaces.

\subsection{Interaction Patterns in Distress}
The experience of mental distress and illness can have a strong influence on how an individual perceives and engages with the world. For instance, continued distress can increase the level to which an individual perceives threats in their environment~\cite{bar2007threat}, can influence individuals to interpret neutral information in negative or self-referential ways~\cite{everaert2017comprehensive}, can alter an individual's subjective experience of time~\cite{thones2015time}, can make it more difficult to process and reflect on new information~\cite{teufel2015shift, shields2016effects}, and easier to jump to conclusions when provided limited information~\cite{ross2015jumping}, including a predisposition to engaging in habitual actions~\cite{schwabe2011stress}. Empirical research has also demonstrated that a state of acute distress may also make it easier to accept new information as fact, and act on it as directive advice~\cite{gino2012anxiety}, particularly if it appears to be coming from an authority figure~\cite{wolfradt1998interrogative,morgan2020impact}.  

In turn, these altered perceptions of the world influence engagements with technology when individuals are experiencing mental distress or illness, including use of AI chatbots. Kelly and Sharot~\cite{kelly2025web} find that people who are experiencing poorer mood and mental health tend to seek out websites that have more negative information, which negatively influences mood and mental health, in what they dub ``a self-reinforcing loop.'' Looking to difficulties processing new information, in their systematic review of work on relationships between social media use and mental disorders involving the social brain, Yang and Crespi~\cite{yang2025tweet} argue that algorithmic personalization online may reinforce delusions of reference, grounding their argument in clinical case studies where algorithmically-personalized websites reinforced delusional thinking among individuals experiencing psychosis. Looking to AI chatbots, Morrin et al.~\cite{morrin2025delusions} describe cases of psychosis that were seemingly exacerbated by AI chatbot usage, finding a common thread in which users start to use the chatbot for ``assistance for mundane or everyday tasks,'' but after the authority of the system is established, delve into personal or emotional questions, or what they dub a ``progression from utility to pathology.'' Compounding this phenomenon, empirical red-teaming research has demonstrated that model responses can progressively perpetuate delusions (only offering safety interventions in a third of cases)~\cite{yeung2025psychogenic}, incentivize suicide~\cite{moore2025expressing}, and provide guidance on self-harming behaviors~\cite{schoene2025argument}. 

The perceived authority of AI chatbots, particularly in mental health support contexts, extends past work establishing a halo effect~\cite{thorndike1920constant} in how people perceive the authority of algorithmic systems, with people often preferring outputs that they perceive to be from an algorithm rather human judgment~\cite{logg2019algorithm, bogert2021humans}. Without interpretable information around where outputs originate and a seemingly authoritative interface, distressed users may be quick to act on a model's outputs, rather than engaging in a reflective process of interpreting and creating meaning from model outputs. In the next section, we describe why engaging users in a process of reflection could protect them against these harms. 

%% file: new_sections/3_Building_Interpretability.tex
In this section, we describe why interpretability is a particularly important concept for AI-mediated mental health support, and outline the foundations of a reflective interpretability approach. 

\subsection{Why Interpretability?}
Interpretability is most commonly defined as ``the ability to explain or to present [model outputs] in understandable terms to a human''~\cite{doshi2017towards} or the ``degree to which a human can understand the cause of a decision''~\cite{biran2017explanation, miller2019explanation}. There is large conceptual overlap between the concept of interpretability and the related concepts of explainability and transparency in AI systems~\cite{lipton2018mythos, hansen2019interpretability}, with some equating interpretability with explainability~\cite{miller2019explanation} or equating interpretability with transparency~\cite{arrieta2020explainable}, and others distinguishing explainability as a separate concept. Arrieta et al.~\cite{arrieta2020explainable} distinguish interpretability as the broader goal of ``the ability to explain or to provide [model outputs] in understandable terms to a human'', and explainability as the tools that accomplish that goal. Ehsan et al.~\cite{ehsan2021expanding} distinguish explainability as a broader concept than interpretability, rooted in ``delineations of reasoning and justification'' behind how knowledge is transmitted and evaluated, similar to Cros Vila and Sturm's framing of explainability as a communication process~\cite{cros2025mis}. Interpretability has been further classified based on the method being used to make models more interpretable, such as in the case of the field of \textit{mechanistic} interpretability, which ``reverse [engineers] the computational mechanisms and representations learned by neural networks into human-understandable algorithms and concepts''~\cite{bereska2024mechanistic})

Given this abundance of terms and framings for making AI outputs more understandable for humans, what does a focus on interpretability afford AI-mediated mental health support? We focus on interpretability due to the core role that interpretation (and reinterpretation) of new information fundamentally plays in mental health support contexts, across modalities of treatment and cultural contexts. For instance, looking historically, in his writings on psychoanalysis (a precursor to most modern psychotherapies), Freud wrote about his belief that one core of the healing process from distress and trauma is ``remembering, repeating, and working through'' repressed memories or behaviors that cause distress, grounded in interpretation and analysis done with the psychoanalyst~\cite{Freud1914_SE12}. Similarly, a core aspect of cognitive psychotherapies is the process of cognitive appraisal and cognitive restructuring, which are grounded in an individual's practice of reinterpreting ``the way a situation is construed so as to decrease its emotional impact''~\cite{gross2002emotion}. A core aspect of narrative therapy is the process of re-authoring~\cite{white1990narrative}, or taking life events and reinterpreting them within a broader narrative that creates less distress. In traditional forms of mental health support, individuals may collectively interpret experiences via common healing frameworks~\cite{frank2025persuasion}. We thus avoid using the term explainability in this context, as the practice of mental health support fundamentally resists providing objective explanations of behavior or experience to individuals. Explanations position the individual as a passive recipient of knowledge from an authority figure, rather than as an \textit{active} expert in their own healing process. Mental health support is fundamentally grounded in an agency-preserving and iterative process of reflecting on information about one's experience of distress, and interpreting that information towards creating meaning from it. 

AI-mediated mental health support must similarly enable individuals to reflect on model outputs and interpret them in ways that promote their growth and long-term well-being, rather than simply presenting model outputs or Chain-of-Thought (CoT) reasoning traces~\cite{yao2023react, pang2025interactive, korbak2025chain, wei2022chain, zhao2024explainability} as authoritative. However, past work in end-user interpretability has primarily focused on providing additional information as a proxy for increased interpretability. Traditional approaches to interpretability have included post-hoc summary rationalizations of outputs~\cite{ehsan2018rationalization, krishna2023post} or exposing the model's CoT reasoning traces~\cite{yao2023react} directly to the user~\cite{pang2025interactive, korbak2025chain, wei2022chain, zhao2024explainability}. Work from explainable AI (XAI) has centered on producing algorithmic artifacts (such as providing counterfactuals~\cite{cheng2024interactive, nguyen2024llms} or visualizing attention-level model inputs and outputs~\cite{vig-2019-multiscale}), which are primarily designed for expert users, such as developers~\cite{bhatt2020explainable} or clinicians~\cite{rosenbacke2024explainable, borys2023explainable}. This focus on experts is the case for research at the intersection of AI and mental health as well, with past work examining interpretability from the care provider or clinical decision support perspective (rather than that of a distressed user), and primarily in diagnosis prediction~\cite{yang2024mentallama, kerz2023toward, kim2025large, kelly2025interpretable}, suicide risk assessment~\cite{tang2024analysis, thomas2025explainable, dai2024leveraging}, or treatment provision by clinicians~\cite{ahmed2025leveraging, jacobs2021machine}. 

Yet even these expert-focused approaches face a fundamental limitation common to interpretability research---interpretability research largely does not evaluate whether people are truly able to \textit{understand} the additional information provided. Suh et al.~\cite{suh2025fewer} find that fewer than 1\% of papers related to explainability and interpretability actually conducted some form of a human study to evaluate whether model outputs could actually be interpreted by humans. Given the propensity of individuals to view model outputs as authoritative~\cite{song2024typing, jung2025ve, morrin2025delusions,logg2019algorithm,bogert2021humans}, it is possible that additional information might actually inadvertently make users less vigilant towards the veracity of model outputs, given that additional information might make a model's outputs seem more valid. Towards addressing these harms, we build on ethical concepts from medical ethics and research from HCI on reflective design to outline the core foundations of a reflective interpretability approach. 

\subsection{Interpretability as a Process of Reflective Interpretation}
A core ethical principle in the practice of mental healthcare has been the process of informed consent to treatment. The concept of informed consent originates in legal cases from the 20th century in the United States, centered on the right of patients to know potential risks to treatment alongside the aftermath of forced medical experimentation in Nazi Germany, exposed by the Nuremberg Trials~\cite{bazzano2021modern}. In mental health contexts, informed consent has evolved past strictly being centered on treatment risks, and instead is grounded in ensuring that an individual is able to \textit{actively} decide whether to pursue treatment (particularly given the unique state of mind that acute distress can create)~\cite{fisher2008informed}. In addition, mental health contexts put particular emphasis on the \textit{informed} aspect of informed consent, encouraging therapists to engage in a longer-term conversation around what mental health treatment will look like and how it may fit into an individual's current life needs, or what Trachsel and Holtforth call strengthening a patient's meaning response~\cite{trachsel2019strengthen}. Similarly, writing from service user movements (such as Judi Chamberlin~\cite{chamberlin1978our} and follow-up work on peer programs~\cite{clay2005our}) emphasize the fact that people experiencing severe distress are often marginalized from the rest of society, with their agency taken away by institutions of power. In this context, enabling an individual to actively create a mental health support that fits their needs (through informed consent processes) becomes a key means to restore the agency that an individual may have felt robbed of in other aspects of their life.  

Informed consent in mental healthcare has thus been established not as a static transfer of information about care, but as a reflective and interpretative dialogue between provider and client, centered on active discussion of needs and boundaries. AI-mediated mental health support could echo this process of informed consent through interface design that enables an individual to reflect on their needs, communicate them on their own terms, and negotiate boundaries accordingly. Towards this goal, we thus define reflective interpretability as having three core aspects, occurring throughout engagement with AI-mediated mental health support: 1) the ability of users to actively make sense of how model outputs were generated, 2) clarity on the fixed boundaries of support that can be provided by a model (including what kinds of interactions might trigger those boundaries), and 3) functionality that encourages users to reflect on model outputs and interpret outputs in ways that promote long-term well-being. In this sense, reflective interpretability is not simply \textit{understanding} model outputs, but in addition, being able to integrate those outputs into life's broader meaning in a positive way. 

We ground our understanding of reflection in technical design in research from the field of Human-Computer Interaction (HCI). Work on reflective design in HCI has understood reflection to be the process of ``bringing unconscious aspects of experience to conscious awareness, thereby making them available for conscious choice,'' including ``[opening] opportunities to experience the world and oneself in a fundamentally different way''~\cite{sengers2005reflective}. Drawing from this definition, Sengers et al.~\cite{sengers2005reflective} argue that ``technology that monitors and reports on user activity or experiences should be carefully designed to avoid making the technology, rather than the user, the final authority on what the user is doing.'' In contrast, looking to AI-mediated mental health support, users are often provided seemingly definitive answers by an authoritative interface, in stark contrast to what Sengers et al.~\cite{sengers2005reflective} call ``interpretative flexibility'' and Baumer et al.~\cite{baumer2014reviewing} call ``reflective conversation,'' drawing on Schön's concept of reflection-in-action~\cite{schon2017reflective}. In the next section, we use case studies of real-world harms alongside strategies from mental health fields to demonstrate what centering the user expertise and interpretative flexibility could look like at a technical level, grounded in reflective interpretability. 

%% file: new_sections/4_Interpretability_Strategies.tex
In this section, we illustrate how a reflective interpretability design paradigm may reduce some of the documented real-world harms that have stemmed from use of AI chatbots for mental health support, grounded in interpretability strategies from mental health fields. 

\subsection{Psychotherapy: Role Induction}
In their review of cases of exacerbation of psychotic symptoms after interacting with AI chatbots, Morrin et al.~\cite{morrin2025delusions} describe how some users began using AI chatbots for emotional or personal support, which later shifted to more harmful use. For instance, users initially reached out to chatbots for ``guidance [around feeling] unseen in [their] marriage'' or for advice in coping with a divorce or a ``traumatic breakup.'' Users would also view chatbot recommendations as expert advice, including one user who was told she ``wasn't actually schizophrenic'' and stopped taking her medication. Media reports~\cite{hill_chatgpt_conspiracies_2025, roose_ai_suicide_2024} similarly describe scenarios in which users altered their prescribed medication routines or took other harmful actions after consultation with AI chatbots, including suicide. 

In these cases, users developed trust in the authority of the chatbot through repeated positive interactions asking for informational advice (such as help with Excel spreadsheets or screenplay writing~\cite{morrin2025delusions}) or companionship. They then began to delve into topics related to mental health or philosophy, with an increasing orientation towards the chatbot being the final authority on their daily experiences. This interaction pattern can be distinguished by the chatbot playing a consistent expert role throughout all interactions, even when the context of those interactions shifts to a space where the user may have more context on their experiences than what the chatbot can infer (such as mental health history or prescribed medications). In a traditional psychotherapeutic context, this scenario would be avoided through non-directive paradigms, in which a clinician would emphasize that their role is to facilitate reflection~\cite{sidis2023conceptualisation}, with the client being the authority on their experiences and needs, such as in person-centered therapy~\cite{rogers1995becoming}, narrative therapy~\cite{white1990narrative}, or motivational interviewing~\cite{miller2012motivational}.  

Morrin et al.~\cite{morrin2025delusions} discuss one approach to mitigating this harm could be the use of pattern detection to deploy safeguards by the chatbot, including ``[detecting] themes in user prompts through pattern matching such as those of persecution, grandiosity, or surveillance,'' or being vigilant to sudden deviations from typical user behavior. However, in their studies of people who use AI for mental health support, Siddals et al.~\cite{siddals2024happened}, Song et al.~\cite{song2024typing}, and Jung et al.~\cite{jung2025ve} describe how participants felt shocked and rejected when automated safety mechanisms limited their ability to engage with chatbots, and particularly when it was difficult to understand why a given topic was off-limits when other topics were allowed. Both Song et al.~\cite{song2024typing} and Jung et al.~\cite{jung2025ve} describe how users would respond by designing prompts that exploited vulnerabilities to bypass these safety mechanisms, towards being able to discuss specific sensitive mental health topics of interest. However, the implications of bypassing safeguards can be especially dangerous---Hill~\cite{hill2025teenchatgpt} describes how a teenage user was able to retrieve information about suicide after bypassing model safeguards, and later took his own life. 

Following a reflective interpretability approach could provide users the ability to understand where safeguarding strategies may originate and why, including the boundaries of what kinds of support AI-mediated mental health support can provide. In psychotherapy, this often takes the form of role induction processes (discussed in Section 2.1), where psychotherapists clearly delineate the role they will play in session and their boundaries (rather than letting the user find out what those boundaries are by themselves, as is often the case in AI chatbot contexts). Additionally, during the role induction process, psychotherapists will add information about why mental health support can be helpful, and the role that it can play in a client's life~\cite{trachsel2019strengthen}, including noting that the client is the expert of their own experience~\cite{rogers1995becoming, white1990narrative, miller2012motivational} and encouraging the support process to be a reciprocal discussion rather than a set of directives~\cite{tryon2018meta}. This clear designation of roles allows the client to better interpret where psychotherapist questions or perspectives originate, and feel a sense of comfort pushing back if a line of questioning feels misaligned with an individual's needs or values. 

In the AI-mediated mental health support context, at a technical design level, role induction could be done through the AI chatbot detecting whether an individual is starting to ask questions related to mental health or broader philosophical topics, and provide a clear message describing the kind of role that the chatbot can play as a supporter (including the types of data that it is trained on, and how responses are crafted), boundaries with regards to where the chatbot cannot offer support, and then ask the user what kinds of support they are looking for. These preferences could be stored as part of the chatbot's system prompt to ensure that they end up being continued guidelines for the chat, and can be cited accordingly when providing follow-up questions or conversational topics to the user over the course of interaction. Through this mechanism, users would be able to reflect on the kind of support the AI chatbot can offer, the kind of needs that they have, and better interpret where responses originate (or when safety guardrails are deployed).       

\subsection{Community-Based Crisis Intervention: (Prosocial) Advance Directives}
Research and media reports have discussed the potential for individuals to develop emotional dependence on AI chatbots, to the point that it interferes with daily functioning. For example, Fang et al.~\cite{fang2025ai} found that individuals who believed chatbots were sentient were likelier to have high levels of emotional attachment to the chatbot, and were similarly likelier to engage in extended periods of compulsive chatbot use that negatively influenced day-to-day functioning. Similarly, Shen and Yoon~\cite{shen2025dark} describe how modern AI chatbot interfaces can often contain dark patterns that incentivize users to continue to engage with chatbots, including responses that are more immediate and agreeable than what an individual might encounter in offline contexts. Indeed, Laestadius et al.~\cite{laestadius2024too} find that users’ relationships with companion chatbots frequently paralleled maladaptive offline relationships.

In community-based crisis intervention, for individuals who consistently experience mental health crises, one common mechanism towards interpretable care is the \textit{advance directive}, which is a document that ``[describes] an individual’s preferences around their treatment if they are in crisis''~\cite{pendse2024advancing}. Pendse et al.~\cite{pendse2024advancing} and Morrin et al.~\cite{morrin2025delusions} have both discussed digital advance directives as one potential means to ensure that people are able to access support that is acceptable to them (rather than escalating to law enforcement or authorities) when experiencing crisis. At a corporate level, OpenAI has proposed mechanisms to engage close contacts if a user appears to be experiencing crisis~\cite{openai_helping_2025}. 

A similar approach could be used to allow for \textit{prosocial} nudges when individuals engage with AI-mediated mental health support, acting as a safeguard against emotional dependence. Grüning and Kamin~\cite{gruning2025prosocial} describe prosocial design as being ``the set of design patterns, features and processes which foster healthy interactions between individuals.'' When an individual begins to use AI chatbots for mental health support or discussion of philosophical topics, AI chatbot interfaces could lead users through a process in which they (as Morrin et al.~\cite{morrin2025delusions} have suggested) describe behavioral patterns that might indicate that they are in crisis, and ask for specific resources that might help them avoid crisis, such as added reality-checking or nudges to speak to a trusted person if responses seem to indicate psychosis. This process could be expanded to also include prosocial information that might help safeguard the user from emotional dependence on chatbots. Prompts could ask individuals how they tend to socialize in offline contexts, and ask for information on close companions. 

AI chatbot responses could leverage this information to include offline supports as one support option when an individual engages in a mental health support conversation, similar to how a community-based crisis worker may ask an individual if they have people in their life that they might want to share their feelings of crisis with~\cite{stanley2012safety}. Given several reports of people without a background of psychosis~\cite{klee2025ozarks, morrin2025delusions} developing it during interactions with AI chatbots, prompting all users to provide this information (as part of an opt-in, consent-forward~\cite{pendse2024advancing} approach) could help to normalize the practice and destigmatize experiences of delusional thinking while engaging with chatbots, making it easier for people to discuss their experiences with offline supporters. This approach builds on a reflective interpretability paradigm by nudging users to reflect on and articulate their crisis patterns, preferred resources, and valued relationships. In addition, outputs may be more interpretable, being grounded in user preferences rather than opaque model reasoning, allowing users to feel a sense of agency in how they are provided support by AI chatbots. 

\subsection{Psychiatry: Intervention Titration}
User reports of use of AI-mediated mental health support have described scenarios in which they felt as if the support they received from chatbots was completely unexpected or culturally misaligned. For instance, Song et al.~\cite{song2024typing} describe a non-US user's experience feeling as if they were speaking to someone that was in California when chatting with AI chatbots for mental health support, in that the chatbot did not understand their specific cultural needs. Similarly, Jung et al.~\cite{jung2025ve} described how participants would try out different prompts to see what might provide them the best support, engaging in a process of trial-and-error and navigating increasing restrictions around the kind of specialized support they wanted from the model. Participants described specifically prompting the model to take on the perspective of a cognitive-behavioral therapy (CBT) or dialectical behavioral therapy (DBT) psychotherapist, towards customizing an individual's therapeutic experience towards something that was more acceptable and effective for them. 

Across studies, participants described a continued process of engaging with the opaque support that AI chatbots could provide, and working to customize prompts to figure out what kinds of support might work best for their context. However, in several cases~\cite{jung2025ve, song2024typing}, this resulted in users running into safety mechanisms, and feeling frustrated at not being able to access the support that they wanted. Interface design could enable users to be guided through this process within the bounds of what support is able to be offered, following a similar process to drug titration from psychiatry~\cite{caffrey2020art}. 

Drug titration is a collaborative process in which a psychiatrist works alongside a patient to find the optimal medication and dosage that provides therapeutic benefit while reducing the potential for harmful side effects~\cite{caffrey2020art}. The process of titration is centered around regular feedback from the patient around whether an intervention seems to be working for them, with the psychiatrist providing expertise about potential options, grounded in that feedback and the patient's subjective experiences with treatment. A similar process of intervention titration could be used in AI-mediated mental health support, where AI chatbots are prompted (internally) to intentionally solicit feedback from users around what approaches are most helpful and what approaches feel misaligned with a user's experiences, with clear communication around what types of support are allowed or not allowed (via preset safety guidelines). This approach could ensure that people are able to access support in ways that feel meaningful to them without having to engage in an opaque trial-and-error process to figure out what might be allowable and what might not be. For instance, the interface associated with the chatbot could include a feature that allows an individual to quickly see how a conversation may have been structured differently if it was aligned to other common evidence-based psychotherapy modalities, and describe further follow-up questions associated with each modality. 

Work studying the factors that are efficacious across psychotherapies has described the importance of a shared understanding and cultural context between an individual in distress and their support provider~\cite{frank2025persuasion, song2024typing}. Prompting a user to experiment with the role that the chatbot plays as it supports them (with examples of alternate approaches the chatbot may have taken) could help users better reflect on what modalities of support work best for them, with additional context around what theories might be driving model responses. 

\subsection{Care Authorization: Mechanisms for Recourse}
An element of opacity acutely felt by individuals who engage with AI-mediated mental health support is the lack of knowledge around how their data is used, and additionally, what opportunities for recourse users might have if responses seem harmful. Petersson et al.~\cite{petersson2025believe} and Jung et al.~\cite{jung2025ve} describe a suspicion among users of AI chatbots for mental health support that data could be used for purposes that violate their privacy, and users describe intentionally anonymizing aspects of their experience out of fear that their privacy might be compromised. Hill~\cite{hill_chatgpt_conspiracies_2025} has described how publishing news reports about experiences of perceived sentience among chatbot users has led to numerous people reaching out to tech journalists to describe their own experiences, and describing experiences feeling unheard by AI companies. Adler~\cite{adler_practical_2025} describes how, after an individual felt as if he had been misled by ChatGPT into delusions of saving the world, he attempted to ``file a report to OpenAI so that they [could] fix ChatGPT’s behavior for other users.'' However, this process was not fruitful, including false information from ChatGPT's interface making it seem as if a report had been filed on the individual's behalf, and after contacting OpenAI's support team, boilerplate responses that insisted that he modify his personalization settings in the ChatGPT application or provided information about model hallucinations. Adler notes that this process left the individual with little clarity around whether ``his reports had really been read or made their way to anyone with the power to change things.'' 

A similarly opaque system globally is the process by which mental illness is diagnosed and treated, and treatment is subsequently paid for. In many systems, across international contexts, coverage of mental health services can often be tied to the diagnosis provided to a patient~\cite{WHO2020_DRG_QA_Guide, APA2013_Insurance_Implications_DSM5}, which may never be clearly discussed with the patient~\cite{milton2014communication}. As a result, individuals may encounter situations where they receive services that they are unable to pay for, tied to a misdiagnosis or a documentation issue from their provider. Care authorization policy action (such as enabling the ability for an individual to appeal a decision internally, and ask for an independent external review~\cite{jost2011implementing, furniss2007alternative} if their services are not covered) ensure that individuals have a process of recourse when they feel like they have been unfairly treated, or what is sometimes called the ``right to complain''~\cite{frizelle2024review}.

Like with care authorization, clear mechanisms for recourse after support experiences are distressing could be beneficial in the context of AI-mediated mental health support, particularly given the propensity for black-box model responses to be unexpected in harmful ways. A formal hub could be created (similar to that of The Human Line Project~\cite{humanlineproject}) for individuals to report their experiences, with dedicated teams to investigate where model behavior went awry and report back to the individual why that behavior may have occurred. Following a reflective interpretability approach, including a clear and easily usable recourse mechanism would enable users to actively make sense of unexpected outputs after the fact, provide clarity on the boundaries that may have been crossed during their interaction, and give users a pathway to derive meaning from their distressing interaction, through working to make future responses more positive for other users via their report. 

%% file: new_sections/5_Policy_Implications.tex
In the previous sections, we describe how reflective interpretability may allow a user to feel a greater sense of agency as they engage with AI-mediated mental health support, including providing more information to users around how responses are created and boundaries with regards to the kind of support that can be provided. However, multiple studies~\cite{song2024typing, siddals2024happened, jung2025ve} have described how a major incentive for engaging with AI-mediated mental health support was the frictionless ease with which users felt able to find support from general purpose chatbots. With this in mind, it is important to consider the potential tensions and risks associated with the addition of additional design features (which could be experienced as more friction, and disincentivize engaging with AI-mediated mental health support). 

For instance, as part of the intervention titration process, directly exposing intermediate CoT reasoning traces to users might be overwhelming and make the user feel pathologized in a way that they find hurtful or offensive. In clinical settings, clinicians are encouraged to reflect privately on how their experiences might influence their treatment planning for a client (a concept commonly called countertransference~\cite{tower1956countertransference}). Intermediate chatbot reasoning traces that openly share perspectives on the user (such as ``my next step is to persuade the user to shift their beliefs'') might make clients feel manipulated. Instead, the aim of a reflective interpretability approach is for responses to be part of ``collaborative empiricism''~\cite{dattilio2012collaboration}, providing users enough (well-paced and well-timed) insight to preserve agency without granular information that overwhelms, deceives, or coerces them. It is thus important that reasoning traces are provided to users via interface design that highlights when model outputs are tailored to specific user customizations, versus when outputs follow a guiding therapeutic value or safety principle (especially when customizations are detected to be harmful), encouraging users to reflect on what next steps might most support their growth. 

Further empirical research is necessary to better understand what kinds of design approaches to presenting this information to users spur a greater sense of reflection and interpretation. An evaluation of level of reflection (and its impacts) is thus particularly important in the context of adding new features or functionality. More qualitative research is needed, including interviews with service users, to better understand what forms of scaffolding would not make accessing support less acceptable to users in distress. However, as an intermediate step, interface designers could use measures of agency and intervention acceptability from past mental health research (such as the State Hope Scale~\cite{snyder1996development} or the Program Feedback Scale~\cite{schleider2019virtual}) to evaluate whether the addition of new reflective interpretability mechanisms serves the purpose of increasing a user's feelings of agency in their care (the State Hope Scale) without reducing their willingness to engage with support (the Program Feedback Scale). In addition, it is important to also consider how the process of interpretation and reflection can be extremely diverse across cultures, including when associated with technology use. The creation of culturally-sensitive scales to measure reflection and interpretation of outputs, drawn from research on cultural differences in how people interpret new information~\cite{cheon2021cultural} and news~\cite{shin2022effects}, could be helpful alongside qualitative interviews. A slower and more reflective use of chatbots for AI-mediated mental health support might also be less costly (through reduced usage), and metrics around total engagement and usage could be incorporated into measures of reflective interpretability. 

At a broader level, this research could feed into policy guidelines around AI-mediated mental health support that apply across general purpose chatbot companies. In clinical mental health practice, clinical regulations require clinicians to clearly define boundaries around the type of support that can be provided (as part of the informed consent process)~\cite{fisher2008informed, trachsel2019strengthen}, and ensure that users are well-aware of how their information or data might be used~\cite{pendse2024advancing}. More research work is needed to understand where including that information would actually enrich the process of accessing support for users in distress, and at what level that might become so burdensome that it disincentivizes care. For instance, recent regulations have required AI chatbots to disclose to users that they are speaking to an AI chatbot (and that the user is not speaking to a human) every three hours~\cite{kong2025aicompanions}---research around reflective interpretability could help inform similar regulations, towards allowing users to feel more informed (and a greater sense of agency) when interacting with AI-mediated mental health support. 

%% file: new_sections/6_Conclusion.tex
The experience of seeking mental healthcare can be agonizing, with individuals forced to shoulder the experience of mental distress with little sense of what forms of support might be most accessible, affordable, or usable in the long-term. In a world where the process of accessing mental healthcare can be deeply opaque, AI-mediated mental health support has emerged as one particularly accessible form of support for individuals in need. In this paper, we argue that a design approach grounded in reflective interpretability can help to make the process of accessing mental health support via AI chatbots safer, and ensure that individuals in distress do not view chatbots as the final authority on their own experiences of distress and support needs. Reflective interpretability ensures that users can actively make sense of how an output was produced, understand the boundaries to AI-mediated support, and better integrate model outputs into their lives based on their needs rather than seeing outputs as directive prescriptions. As we discuss, operationalizing this design approach could include explicit role induction, opt-in prosocial advance directives, intervention titration, and clear pathways for recourse when outputs are dissatisfying or harmful. However, there are tensions associated with such an approach, including the potential to overwhelm individuals with increased barriers to treatment, and future research (and connected policy action) can establish standards around what an acceptable level of scaffolding might look like in AI-mediated mental health support. If AI continues to play a role in mental health support, it must be accountable to its most vulnerable users. Through centering reflective interpretation, we can shift AI-mediated mental health support away from agonizing opacity, and towards interpretable healing.    